\documentstyle[aps,epsf,prl,multicol]{revtex}
\input{epsf}

\begin{document}

\title
{Director precession and nonlinear waves in nematic liquid crystals 
under elliptic shear}

\author{T. B\"orzs\"onyi$^{1,*}$ \'A. Buka$^{1}$ A. P. Krekhov$^{2,3}$ 
O. A. Scaldin$^3$ and L. Kramer$^2$}
\address{$^1$Research Institute for Solid State Physics and Optics,
  Hungarian Academy of Sciences,  H-1525 Budapest, P.O.B.49, Hungary\\
  $^2$Physikalisches Institut, Universit\"at Bayreuth, D-95440 Bayreuth, 
  Germany\\
  $^3$Institute of Molecule and Crystal Physics, Russian Academy of Sciences,
 450025 Ufa, Russia\\
 $^*$ Present address: Groupe de Physique des Solides, CNRS UMR 75-88, 
Universit\'es Paris VI et VII, Tour 23, 2 place Jussieu, 75251 Paris Cedex 05, 
France\\
} 
\maketitle

\begin{abstract}
Elliptic shear applied to a homeotropically oriented 
nematic above the electric bend Fr\'eedericks transition (FT) 
generates slow precession of the director. 
The character of the accompanying nonlinear waves changes 
from diffusive phase waves to dispersive ones exhibiting  spirals and 
spatio-temporal chaos as the FT is approached from above.
An exact solution of the flow alignment equations captures the observed
precession and predicts its reversal for non-flow aligning materials.
The FT transforms into a Hopf bifurcation opening the 
way to understand the wave phenomena.

\end{abstract}
\pacs{PACS numbers: 61.30.Gd, 47.20.-k, 47.20.Ky, 47.20.Lz}

\begin{multicols}{2}
\narrowtext


Liquid crystals (LCs) exhibit a multitude of interesting nonlinear dynamical 
phenomena, like electrically and thermally driven convection, 
flow-induced and optical instabilities \cite{optic,LCbook}.
Nonlinear dispersive waves as observed in particular in
excitable and oscillatory
chemical reactions \cite{BZ} are rarely seen in LCs. An exception
are the 2D waves that modulate the oscillating bimodal pattern found
in electroconvection of nematics \cite{sano1} for which 
only a phenomenological description has been developed.

Actually, nonlinear dispersive waves are obtained in a much simpler situation 
when a local precession of the 
director is generated by an elliptic shear flow in a
homeotropically anchored, Fr\'eedericks distorted nematic slab. 
As will be shown here, the dynamics can then be nearly conservative, 
in contrast to the well studied situation where the 
director is set into motion by a rotating magnetic or electric field 
\cite{frisch1} and  the dynamic phenomena are limited by the strong dissipation. 
In the presence of elliptic shear the FT transforms into a Hopf bifurcation 
and the observed transition to chaos can possibly be related to the 
Benjamin-Feir instability, see e.g. \cite{CrHo93}.

Elliptic shear flow in homeotropically aligned nematics has been studied 
intensively in the past in view of an instability leading to
roll patterns \cite{pigu77,edvpmbook}.
For that purpose rather large shear angles are needed and no external 
field is applied.
The only previous study of the Fr\'eedericks distorted state in the presence of
elliptic shear was, to our knowledge, carried out by Dreyfus and Pieranski.
They observed some of the phenomena discussed below
without a conclusive interpretation \cite{drpi81}. 


In this work the nematic layer (thickness $d$) has been sandwiched 
between clean, $SnO_2$ coated 
float glass plates ($xy$ plane) resulting in homeotropic anchoring 
(nematic director ${\bf \hat n}$ perpendicular 
to the bounding plates) of the substance Phase 5A (Merck). 
A voltage $U$ applied across the sample induces at a critical value $U_F$
a bend Fr\'eedericks transition (FT) 
(the dielectric anisotropy $\varepsilon_a$ of Phase 5A is negative ).
Above $U_F$ the director tilts away from the $z$ direction.
The tilt angle $\theta$ can be controlled by varying $U$.
The degeneracy with respect to the azimuthal angle $\phi$ 
leads to the well known
"schlieren texture" seen between crossed polars  \cite{degennes}.
It contains defects ({\it umbilics}) with topological charge 
$\pm 1$ depending on whether the in-plane director makes a $2\pi$ or $-2\pi$ 
rotation on a closed loop around them. 
These can be distinguished by rotating the crossed polars.

Elliptic shear flow has been generated by oscillating the bottom glass
plate in the $x$ direction [$x(t) = A_x \ \cos(\omega t)$]
and the top plate along $y$ [$y(t) = A_y \ \cos(\omega t + \Phi)$;
unless stated otherwise the phase shift will be $\Phi=-\pi/2$].
The oscillations were produced by loud speakers.
The precession of the director was observed as the local oscillation of 
the transmitted light intensity that results in wave-like propagation of 
bright and dark domains.
The recordings were studied by digital image analysis.
Details of the experimental setup are described in \cite{bbkk,Bophd}.

\begin{figure}[h]
\parbox{9cm}{
\begin{center}
\epsfxsize=8.605cm
\hspace*{-1.096cm}
\epsfbox{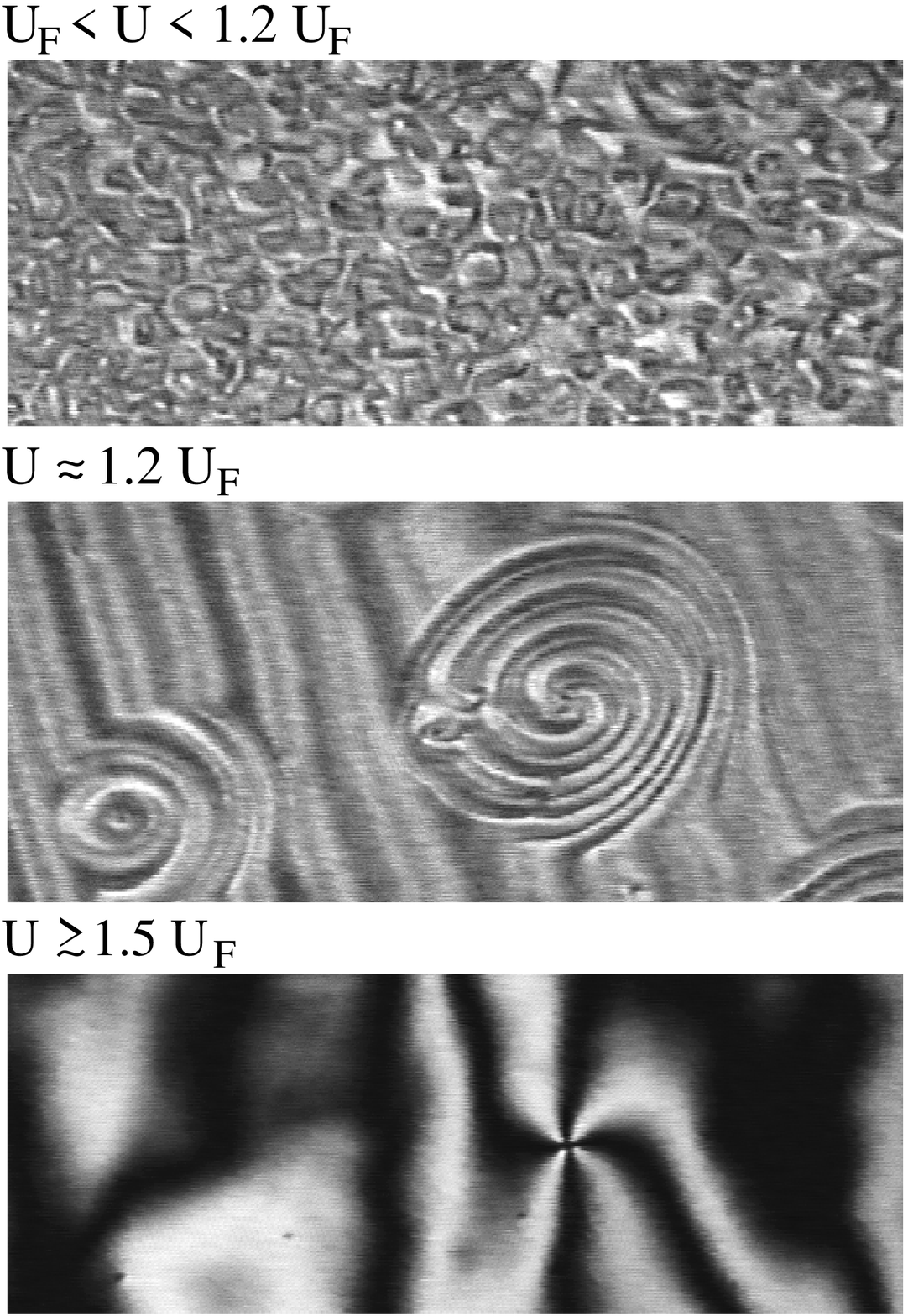}
\end{center}
}
\vspace*{-0.25cm} 
\caption{
Snapshots in the different regimes (crossed polars).
Circular shear, $A_x=A_y=3.4$ $\mu$m, $f=155$ Hz, and $T=28.2 ^o$C.
\label{regimes} }
\end{figure}

At sufficiently high voltages the director 
orientation varies slowly in space and precesses almost homogeneously 
in time (Fig.\ref{regimes} bottom). 
Around $1.2 U_F$  inhomogeneities
emit traveling waves and umbilics generate spiral waves, 
very similar to those observed in oscillatory and excitable chemical
reactions \cite{BZ}.
The longer waves in the background originate from the lateral cell boundaries
(Fig.\ref{regimes} middle).
At lower voltages spiral pairs seem to be created spontaneously (without
umbilics) and one observes spatio-temporal chaos (Fig.\ref{regimes} top).

\begin{figure}[h]
\parbox{9cm}{
\begin{center}
\epsfxsize=7.05cm
\hspace*{-1.06cm}
\epsfbox{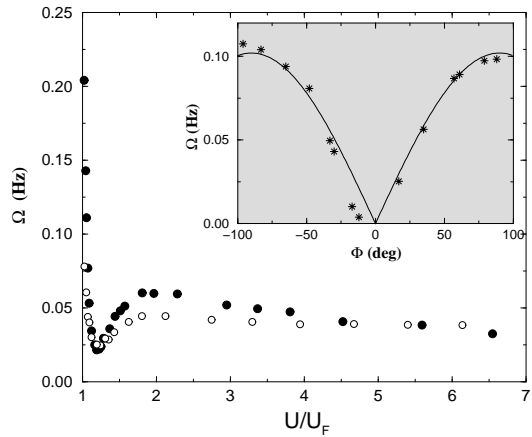}
\end{center}
}
\vspace*{0.cm} 
\caption{
The precession frequency $\Omega$ as a function of the applied voltage 
for $ T = 29 ^o$C,  $U_F = 8.4$ V ($\bullet$)
and   $ T = 53 ^o$C, $U_F = 10.7$ V ($\circ$).
The other parameters were: $d = 20$ $\mu$m,
$A_x = 3.1$ $\mu$m,  $A_y = 3.5$ $\mu$m, $f=155$ Hz and $\Phi  = 80 ^o$.
In the insert we show $\Omega$ as a function of the phase shift $\Phi$
for $A_x=A_y=4$ $\mu$m, $f=122$ Hz, $d=20$ $\mu$m,  $U/U_F = 2.3$ and
$T = 24.5 ^o$C (*). 
The continuous line shows $\Omega = 0.102 |\sin\Phi|$. 
\label{loomfut} }
\end{figure}
\begin{figure}[h]
\parbox{9cm}{
\begin{center}
\vspace*{-1.25cm} 
\epsfxsize=8.725cm
\hspace*{-1.6cm}
\epsfbox{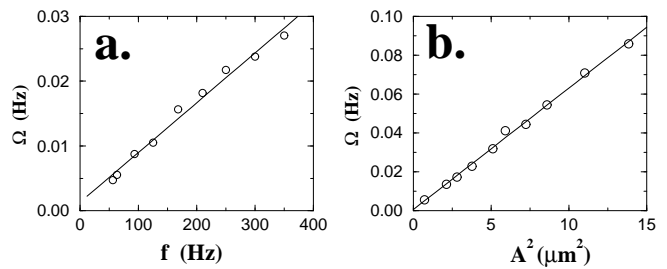}
\end{center}
}
\vspace*{0cm}
\caption{
Precession frequency vs. driving frequency and amplitude.
The parameters are: {\bf a.)} $U/U_F = 2.3$,  $d = 20$ $\mu$m,
$A_x = 0.8$ $\mu$m,  $A_y = 1$ $\mu$m and $ T = 25.5 ^o$C.
{\bf b.)} 
Circular shear,  $d = 20$ $\mu$m,  $f = 122$ Hz,  $ T = 28.5 ^o$C and  
$U/ U_F = 2.3$.
\label{loomf} }
\end{figure}

In Fig.\ref{loomfut} the dependence of the precession frequency $\Omega$
on the voltage is shown for two temperatures.
Clearly $\Omega$ behaves differently in the different regimes 
(see Fig.\ref{regimes}).
The dependence of $\Omega$ on the frequency and amplitude of the applied 
shear is given in Fig. \ref{loomf}. The linear dependencies can 
be understood  from symmetry arguments and dimensional analysis as shown in the following.


In order to describe the director
precession we look for solutions of the nematodynamic equations
where the director ${\bf \hat n}$ and the velocity ${\bf v}=(v_x,v_y,0)$
depend only on the coordinate $z$ and time $t$.
Measuring lengths in units of $d$, times in units of $1/\omega$, 
we introduce
\begin{eqnarray} \label{dimvar}
        k_i = K_{ii} / K_{33} \; , 
        \lambda=\alpha_3/\alpha_2 \; ,
        \tau_d = \gamma_1 d^2 / K_{33} \; ,
	\gamma_1=\alpha_3 - \alpha_2 
        \nonumber \\
        \epsilon = 1/(\tau_d \omega) \; ,
        e^2 = U^2/U_F^2 \; ,
        U_F=\sqrt{K_{33}\pi^2/(\varepsilon_0|\varepsilon_a|)} \;  \ \
\end{eqnarray}
with $K_{ii}$ the elastic constants and $\alpha_i$ the Leslie coefficients.
Setting 
%
$N = n_x + i n_y \; , \ \ V = v_x + i v_y$,
%
one can write the nematodynamic equations concisely in complex notation.
The torque balance equation \cite{edvpmbook,degennes} is
\begin{eqnarray}   \label{eqN}
        N_{,t} =
        \frac{n_z}{(1-\lambda)}\left( V_{,z} - \frac{1+\lambda}{2}
        N ( N \overline{V}_{,z} + \overline{N} V_{,z} )\right)
        \nonumber\\
        - \epsilon \left\{
         k_1 n_z n_{z,zz} N
        - [N_{,zz} - N ( N \overline{N}_{,zz} +
                  \overline{N} N_{,zz} )/2]
        \right. \nonumber\\ \left.
        + (1-k_2) [ N ( B^2 + i B_{,z} )/2 + i N_{,z} B ]
        - \pi^2 e^2 n_z^2 N  \right\} \; ,
\end{eqnarray}
with
$ n_z = \sqrt{1-|N|^2} $,
$ B = i(N \overline{N}_{,z} - \overline{N} N_{,z}) $,
and $_{,i}$ denote derivatives.

For frequencies such that the viscous penetration depth satisfies
$l = \sqrt{\gamma_1/\rho\omega}>>d$ [$f=\omega/(2\pi)$ below about $40$
kHz for $d=20$ $\mu$m], and small dimensionless shear amplitudes
$a=A_x/d$ one has a linear flow field, i.e.
$V,_z=a(\cos{t}+i b\sin{t})$ where $b=A_y/A_x$ is a measure of
ellipticity ($b=1$ for circular shear).

It is useful to first neglect any space dependence, which is a good 
approximation when the Ericksen number $a/\epsilon$ \cite{edvpmbook} is large.
Consistently, one then also has to discard the electric field, so that
in Eq.(\ref{eqN}) one is left with the terms in the first line.
Introducing angles by writing $N=\sin{\theta} \exp{i \phi}$ one can rewrite the equation  as
 \begin{eqnarray}   \label{ODEs}
  \theta,_t&=&a' (\cos^2{\theta} - \lambda \sin^2{\theta})
   [\cos{t} \cos{\phi} + b \sin{t} sin{\phi}] \; ,
    \nonumber\\
  \phi,_t&=&a' \cot{\theta} [-\cos{t} \sin{\phi} + b \sin{t} \cos{\phi}]\; ,
 \end{eqnarray}
where $a'=a/(1-\lambda)$. 
For rectilinear shear ($b=0$) one recovers the flow-aligned
solution $\cot^2{\theta}=\lambda,\ \phi=0$. 

Equations (\ref{ODEs}) represent a conservative, reversible dynamical 
system.
For $\lambda =0$ the director is advected passively by the velocity field
and Eqs.  (\ref{ODEs}) separate into
 \begin{eqnarray}   \label{zero}
  \partial_t(\tan{\theta} \cos{\phi})=a \cos{t}\; ,
  \; \partial_t(\tan{\theta} \sin{\phi})=ab \sin{t}\;.
 \end{eqnarray}
The solutions for arbitrary initial conditions can easily be written.
They describe simple, closed $2 \pi-$periodic orbits, which either include 
or exclude the origin (homeotropic orientation). Clearly this case can be 
generalised to arbitrary time dependence of the flow.

\begin{figure}[h]
	\parbox{8cm}{
	\begin{center}
        \vspace*{-1.25cm} 
	\epsfxsize=6.05cm
        \hspace*{-0.26cm}
	\epsfbox{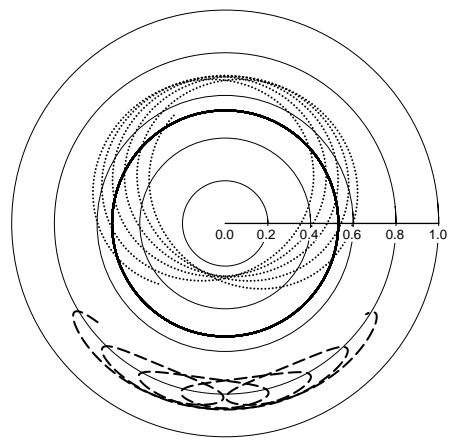}
	\end{center}
	}
	\vspace*{-0.05cm}
\caption{Three orbits of the in-plane director under circular flow 
for $\lambda=0.2$, $a=0.5$. 
The thick circle represents a pure rapid rotation around 
the $z$ axis relevant below the FT. 
(In the elliptic case this orbit transforms into an approximate ellipse.) 
Other orbits exhibit the slow precession. 
We show examples with a small average tilt (dotted), expected to be relevant
 slightly above the FT, and with a large tilt (dashed).
	\label{ODE} }
\end{figure}

For circular flow ($b=1$) Eqs.(\ref{ODEs}) are integrable even for
$\lambda \ne 0$.
Then the terms in square brackets become $\cos{(\phi-t)}$ and $-\sin{(\phi-t)}$,
respectively. Introducing the phase lag $\varphi=\phi-t$ the equations become
autonomous. Transforming them into the 2nd-order ODE 
 \begin{eqnarray}   \label{2nd}
  \varphi,_{tt}=(\varphi,_t+1)(2\varphi,_t+1) \cot{\varphi} 
  -\lambda a'^2 sin{\varphi}\cos{\varphi}
 \end{eqnarray}
one can verify that the quantity
 \begin{eqnarray}   \label{constant}
   C=\frac{(\varphi,_t +1 -\lambda a'^2 \sin^2{\varphi})^2}
   {(2\varphi,_t +1 - \lambda a'^2 \sin^2{\varphi})\sin^2{\varphi}}
  \end{eqnarray}
is a constant of motion. 
Solving for $\varphi,_t$ one obtains the period $T$
of the motion as an integral which can be solved analytically giving
 \begin{equation}   \label{period}
   T=\int_{0}^{2 \pi} d\varphi / \varphi,_t=2 \pi /\sqrt{1- \lambda a'^2}
 \end{equation}
(independent of $C$!). For $\lambda \ne 0$ the
orbits are in general quasi-periodic. Thus, in addition to the rapid 
oscillation, the director performs a slow precession with frequency
$\Omega=(1-2 \pi/T)\omega \approx (\lambda/2) a'^2\omega$ in physical
units. The precession is for flow-aligning materials positive (same sense 
of rotation as the elliptic shear)  and negative otherwise.
In Fig. \ref{ODE} some typical orbits are shown.


Simulations show that for  $b \ne 1$ (not too small) and $\lambda \ne 0$ the 
slow precession persists and has the same sign as for $b=1$. 
One has $\Omega\approx (\lambda/2) a'^2 b \omega$, which can also
be derived analytically for small shear flow amplitudes.

The dependence of $\Omega$ on the ellipticity is explored experimentally
most easily by keeping $A_x=A_y$ fixed and changing the phase shift $\Phi$.
This driving is actually equivalent to one with phase shift $-\pi/2$
and $A_x'=\sqrt{2} A_x \cos(\Phi/2),\ A_y'=\sqrt{2} A_y \sin(\Phi/2)$, 
as can be seen by transforming into a rotated coordinate system.
Thus $a^2 b=(A_x A_y/d^2) |\sin{\Phi}|$.
In Fig. \ref{loomfut} (insert) the experimental results are compared
with the theoretical prediction.
$\Omega/\omega$ was always found to be positive as expected for the 
flow-aligning case.
From the magnitude of $\Omega$ at $U/U_F \approx 2$ and
$T=25 \dots 30$$^o C$ in  Figs. 2, 3a, and 3b.
we find $\lambda=0.008 \pm 0.004$, which appears consistent with the very
small values found in \cite{GrKnSchn92} and falls within the 
experimental uncertainty of those measurements.

Now we discuss the effect of the terms in curly brackets in Eq. (\ref{eqN}).  
These terms represent singular perturbations. They introduce dissipation into 
the system, which, together with the boundary conditions, produces the 
attractors. For $b=1$, $a<<1$ and at lowest order (in $\epsilon$), this reduces 
to a (z-dependent)    selection from the family of solutions of the 
conservative system.  Although this approximation breaks down near the bounding 
plates it can give useful results (see the rectilinear case \cite{KKBC93}). One 
then expects that (aside from the slow precession) the director performs 
small-amplitude oscillations around the director distribution   $\theta_0(z)$ 
obtained from equating the curly brackets in Eq.(\ref{eqN}) to zero (with 
homeotropic boundary conditions $N=0$), which  corresponds to the usual 
(static) bend Fr\'eedericks distortion.  In this approximation the frequency of 
the slow precession should be as calculated above.  We also analysed slow 
modulations in the $xy$ plane and found them to  give rise to diffusive phase 
waves as observed well above $U_F$.

The experimental results indicate that corrections to the above behavior
are important, in particular near the FT.
We have analysed corrections only in limiting cases and found the contribution 
to the slow precession to be positive. 
In particular, one should note that in the presence of circular shear 
{\em the FT is transformed into a Hopf bifurcation}, as can be seen purely 
from symmetry arguments:
below the transition the director rotates symmetrically around the 
homeotropic orientation.
At the FT this symmetry is broken and generically a new frequency develops 
(no chiral symmetry in the rotating frame) which leads 
to the precession.

Consequently, near the FT a description in terms of a complex 
Ginzburg-Landau equation should be possible.
For low frequencies ($\epsilon \sim 1$)
and near the FT we have performed a fairly full analysis.
We find the FT to be slightly delayed and  modulations in the $xy$ plane 
to exhibit strong  dispersion, which presumably leads to the observed
modulational (Benjamin-Feir) instability.
For $1 > b > b_c \approx 0.5$ the behavior remains qualitatively
similar.
At $b_c$ there is a crossover {\em from oscillatory to excitable} behavior of 
the system \cite{excite}.

In summary: the theory describes the slow precession, at large fields
essentially quantitatively. The scenario of the waves on the
background of the slow precession (diffusive phase waves at large 
fields changing to amplitude waves with dispersion and eventually 
Benjamin-Feir chaos at low fields) can be understood qualitatively. 
There remains to be done a quantitative analysis at low fields as well 
as an experimental test
of the most provocative prediction, namely the reversal of the slow
precession for non-flow aligning materials in situations where the
elasticity-induced effects are small.

We have observed similar behavior in the nematics MBBA and DOBHOP
($\varepsilon_a < 0$). Also, in 5CB ($\varepsilon_a > 0$)
we generated the director precession by inducing the
director tilt by surface effects: the bottom plate of 
the horizontal cell was heated  above the phase 
 transition point ($34.5^o$ C) so that a nematic-isotropic interface developed 
in the $xy$ plane. The surface energy of this 
interface is minimized when the director encloses an angle $\sim 64^o$ 
with the surface normal \cite{faetti90}.

We have also performed experiments with linear mechanical vibration along $z$ 
(compression) generated by piezo crystals ($f \approx 5$ - $100$ kHz) 
attached to one of the bounding plates. 
This presumably induces Poiseuille  flow.
As before, the slow precession occurred only in the 
Fr\'eedericks distorted state.
Here the phase waves are typically emitted from certain 
locations in the form of target patterns, which presumably result from spatial 
inhomogeneities in the flow. The waves behave diffusively (even near 
$U_F$), which is probably due to the fact that at the high
frequencies used the elastic contributions to the precession are irrelevant
\cite{reversal}. For more details, see \cite{Bophd}.

Director precession and phase waves have been observed previously in cells that 
were excited piezo electrically at frequencies around $50$ kHz \cite{chu82}.
The piezo crystal formed one of the bounding plates. 
Phenomena reminiscent of the phase waves were also seen in 
planar and homeotropic cells without electric field at frequencies 
$10$ kHz $< f <$ 1 MHz \cite{scudi}.
There the waves originated from orientational defects at the surface.
Although in these cases the precise form of the excitation was not clear we 
suggest that the mechanism presented here forms the basis of the phenomena.

Finally we note that the precession should be observable 
also in smectic C layers excited elliptically.  
Possibly the effect can be used to measure the analog of $\alpha_3/\alpha_2$.

We wish to thank P. Coullet, P. Pieranski, T. T\'el and M. Khazimullin for 
illuminating discussions.
The hospitality of the Max-Planck-Institute for the Physics of 
Complex Systems at Dresden, where  part of  the work was performed, is 
greatly appreciated. 
Financial support by OTKA-T014957, EU (TMR ERBFMRXCT960085) DFG 
 (Grant Kr 690/12-1), INTAS (Grant 96-498), and Volkswagen-Stiftung (I/72 920)
 is gratefully acknowledged.


\begin{thebibliography}{10}

\bibitem{optic}
F.~Simoni {\em Nonlinear optical properties of liquid crystals}
(World Scientific, Singapore, 1997); G. Demeter and L. Kramer \newblock 
{\em Phys. Rev. Lett.}, in press.

\bibitem{LCbook}
See e.g. \newblock {\em Pattern Formation in Liquid Crystals}, 
A.~Buka and L.~Kramer, editors (Springer-Verlag, New York, 1996).

\bibitem{BZ}
See e.g. A.~de~Wit, \newblock {\em Adv. Chem. Physics}, I. Prigogine and 
S.A. Rice, editors (John-Wiley, New York, 1999) Vol.109, p. 435.

\bibitem{sano1}
M.~Sano, H.~Kokubo, B.~Janiaud, and K.~Sato,
\newblock {\em Progresses in Theoretical Physics}, {\bf 90}, 1 (1993);
V.~A. Delev, O.~A. Scaldin, and A.~N. Chuvyrov,
\newblock {\em Mol. Cryst. Liq. Cryst.}, {\bf 215}, 179--186 (1992).

\bibitem{frisch1}
C.~Zheng and R.B.~Meyer, \newblock {\em Phys. Rev. E.}, {\bf 56}, 5553 (1997); T.~Frisch, S.~Rica, P.~Coullet, and J.~Gilli,
\newblock {\em Phys. Rev. Lett.}, {\bf 72}, 1471 (1994);
S. Nasuno and N. Yoshimo and S. Kai, \newblock {\em Phys. Rev. E.}, {\bf 51}, 1598 (1995).

\bibitem{CrHo93}
M.~C. Cross and P.~C. Hohenberg, \newblock {\em Rev. Mod.} Phys. {\bf 65}, 
851 (1993).

\bibitem{pigu77}
P.~Pieranski and E.~Guyon,
\newblock {\em Phys. Rev. Lett.}, {\bf 39}, 1281 (1977);
E.~Dubois-Violette and F.~Rothen,
\newblock {\em J. Phys. France}, {\bf 39}, 1039 (1978);

\bibitem{edvpmbook}
E.~Dubois-Violette and P. Manneville, in \cite{LCbook}.

\bibitem{drpi81}
J.-M. Dreyfus and P.~Pieranski.
\newblock {\em J. Physique}, {\bf 42}, 459 (1981).

\bibitem{degennes}
P.~G. de~Gennes and J.~Prost,
\newblock {\em The Physics of Liquid Crystals}. Clarendon Press, Oxford,
1993;
S.~Chandrasekhar,
\newblock {\em Liquid Crystals}. Dover Publication, New York, 1981.

\bibitem{bbkk}
T.~B\"orzs\"onyi, \'A.~Buka, A.P.~Krekhov, and L.~Kramer,
\newblock {\em Phys. Rev. E.}, {\bf 58}, 7419 (1998).

\bibitem{Bophd}
T. B\"orzs\"onyi, 
doctoral dissertation,  Budapest, 1998 (in English).

\bibitem{GrKnSchn92}
H.~H. Graf, H.~Kneppe, and Schneider, Molecular Physics {\bf 77}, 521 (1992).

\bibitem{KKBC93}
A.P.~Krekhov, L.~Kramer, J. Physique II, {\bf 4}, 677 (1994),
{\em Phys. Rev. E.}, {\bf 53}, 4925 (1996).

\bibitem{excite}
S.C.~M\"uller, P.~Coullet, and D.~Walgraef, \newblock {\em Chaos}, {\bf 4},
443 (1994).


\bibitem{faetti90}
S.~Faetti, 
\newblock {\em Mol. Cryst. Liq. Cryst.}, {\bf 179}, 217 (1990).

\bibitem{reversal}
In the immediate neighborhood of the FT and at not too high temperatures, a 
reversal of the precession was observed in Phase 5 and MBBA.

\bibitem{chu82}
A.~N.~Chuvyrov,
\newblock {\em Zh. Eksp. Teor. Fiz.}, {\bf 82}, 761 (1982) [Sov. Phys. JETP
  {\bf 55}, 451 (1982)];
A.~N.~Chuvyrov, O.~A. Scaldin, and V.~A. Delev.
\newblock {\em Mol. Cryst. Liq. Cryst.}, {\bf 215}, 187 (1992).

\bibitem{scudi}
F.~Scudieri,
\newblock {\em Ann. Phys.}, {\bf 3}, 311 (1978).

\end{thebibliography}



\end{multicols}

\end{document}